# Current states of a doubly connected superconductor with two point contacts

V. P. Koverya, A. V. Krevsun, S. I. Bondarenko, and N. M. Levchenko

*B. Verkin Institute for Low Temperature Physics and Engineering of the National Academy of Sciences of Ukraine, prospekt Lenina 47, Kharkov 61103, Ukraine*

The distribution of the transport current in branches of a doubly connected superconductor (DCS) with two clamping point contacts is investigated experimentally when the contacts are in the different states: (a) one or both contacts are in the critical state, (b) both are in the supercritical (resistive) state. In the state (a) the transport current frozen in the DCS and injected through one of the contacts can be increased or decreased by tuning the value and polarity of the current passing through the other contact. In the state (b), the current self-oscillations the amplitude and the frequency of which depend on the value of the injected transport current appear in the branches of the DCS. A role of the parametric Josephson inductance and resistivity of the contact in formation of its critical state and in distribution of the current in the branches of the DCS is discussed.

## INTRODUCTION

Doubly connected superconductors (DCSs) are the most-used superconducting devices. Among these are superconducting quantum interference devices (SQUIDs) as detectors of a superweak magnetic field of infralow frequency; superconducting short-circuit coils as magnetic field sources not consuming energy; superconducting magnetic shields in form of hollow cylinders. Doubly connected superconductors are also parts of more complicated structures including nanostructures of high-temperature ceramic superconductors and clamping point contacts, in particular, appearing in multi-cored superconducting cables. So to investigate physical properties of various types of DCSs is topical. The simplest structure of DCSs is a superconductor ring or a hollow cylinder. Previously DCSs have been investigated either in form of continuous rings with macroscopic dimensions (the diameter and thickness of a wall is much larger than the magnetic-field penetration depth $\lambda$ and the coherence length $\xi$) (Ref. 1) or in form of microscopic rings and cylinders with small inductance $L_0$, satisfying the condition $\Phi_0^2/2L_0 > \kappa T$ ($\Phi_0$ is the magnetic-flux quantum, $T$ is the ring temperature, $\kappa$ is the Boltzmann constant), containing[2] or not (in this case the wall thickness is less than $\lambda$) (Ref. 3) different types of the Josephson contact. The subject of our investigations is DCSs of the other type. Using the nomenclature of a ring, we have investigated rings containing a given type of a weak coupling (not necessarily of the Josephson type), for which the inverse energy relation ($\Phi_0^2/2L_0 < \kappa T$) is fulfilled without a limitation on the thickness of a ring wall. In particular, we have already found new nontrivial properties of such DCSs related to the current distribution in inductively asymmetric branches of DCSs with a local segment in one of them with reduced critical current,[4] and also in DCSs one branch of which is a clamping point contact (PC).[5]

The aim of the present work is investigation of the current distribution in branches of DCSs similar to that described in Ref. 5 and containing not one but two PCs, one contact each for the branches. The locally injected transport current can be either equal to or higher than the critical current of contacts. In this case, as it is shown below, there appear phenomena which are not realized in DCSs with one PC.

## EXPERIMENTAL PROCEDURE

The diagram of the investigated DCS is shown in Fig. 1. The structure was made of either two niobium or niobium and tantalum microwires with the diameter of 0.07 mm.

One of them together with the shunts 2–3 comprises a coil with a few turns of a microwire (with the inductance $L = 5 \cdot 10^{-6}$ H) enveloping a sensor of a fluxgate magnetometer (FM). The other (with the shunts 1–4) is placed on a dielectric substrate perpendicularly to the first one and is pressed down to it mechanically at two points of their intersection to create two PCs. Thus the DCS circuit consists of two asymmetric in length and critical current of branches, one of which has the length of a PC and is comparable with the coherence length of the used superconductors and the other long branch of a few centimeters has an inductance approximately equal to the inductance $L$ of the mentioned coil. The constant transport current $I_t$ was injected into the DCS either through the PC1 with the shunts 1–2 or through the PC2 with the shunts 3–4, or otherwise at the same time through both PCs from different current sources. The used method of the current injection into the DCS can be considered as local in contrast to the traditional nonlocal used, in particular, by us in Ref. 4 when the distance between shunts is significantly larger than the coherence length of superconductors. As became apparent from comparison of results of our works,[4,5] the method of current injection into the structure plays an important role and affects the current distribution in a DCS. The magnetometer is served for contact-less measuring the magnetic field of the current flowing through the coil. The relation between the current and the magnetic field is found in advance (before pressing the microwires) by using the known current applied to the leads 2–3. This allows to determine the current value in the branch of the DCS. To clarify features of the transport current distribution in the branches of this DCS a dependence of the current $I_L$ in the branch with the coil on the transport currents injected in one or other PC were registered. In what follows, in order to distinguish the transport current injected through the shunts 1–2 from that injected through the shunts 3–4, these currents are denoted as $I_{t1}$ and $I_{t2}$. Altogether ten samples of the DCS were investigated. During the measurements they were located in liquid helium with the temperature 4.2 K. A cryogenic part of the measuring set-up was surrounded by a ferromagnetic shield to reduce an influence of electromagnetic noise on measurements results. In particular, the magnetic field of the commercial frequency (50 Hz) was reduced by the factor of 200.

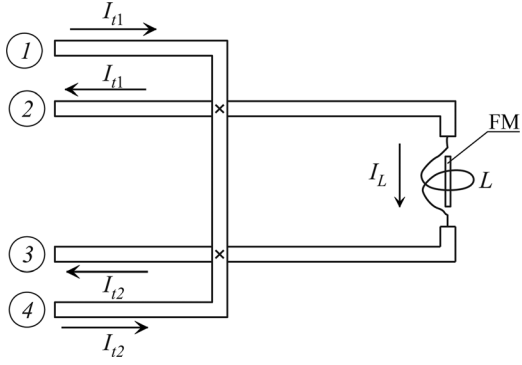

FIG. 1. A circuit of doubly connected superconductor with two clamping point contacts. Their location corresponds to points of a mechanical contact between two microwires where they intersect each other and is indicated by a sign ×, FM is a sensor of a fluxgate magnetometer, 1, 2, 3, 4 — shunts.

**EXPERIMENTAL RESULTS AND THEIR DISCUSSION**

First of all we will consider the transport current injection into the DCS through one of the PCs. Fig. 2 shows three main types of dependences $I_L(I_{t1})$ observed in different samples of the investigated DCS. They differ in the ratio between the height of a hysteresis part along the $I_L$ axis and the height of current steps. In the dependence in Fig. 2(a) this height is larger than heights of steps. In Fig. 2(b) these values are equal, and in Fig. 2(c) the height of the step exceeds that of the hysteresis part. As seen from the dependences, this distinction affects the form of their hysteresis-free parts.

Common features of the dependences are the presence of the hysteresis region 1–2–3–4–5–6 with the equidistant current steps in the regions 6–1–2 and 3–4–5 and the nonlinear hysteresis-free regions 2–7 and 5–8 which correspond to the transition into the resistive state of both PCs. The width of the hysteresis regions along the horizontal axis is equal to the doubled values of the critical current of the PC ($I_{c1}$) through which the transport current $I_{t1}$ is injected into the DCS. The height along the vertical axis is equal to the doubled value of the critical current of the second PC ($I_{c1}$). In this case, in some regions of the mentioned hysteresis-free nonlinear parts of the dependences $I_L(I_{t1})$ at a fixed value of the current $I_{t1}$, periodic self-oscillations (SOs) of the current $I_L$, which vary in amplitude $\Delta I$ and frequency, are observed. The parts with the SOs of the current are shown in Fig. 2 by vertical arrows. The inset in Fig. 2(a) shows one typical example of SOs of the current registered by means of an electronic oscilloscope (top) and an electromechanical recorder with the time constant of 0.15 s. For the mentioned part of the record the frequency of SOs is 4 Hz. As a rule, SOs with a higher frequency have a lower amplitude. We observed SOs with the frequencies from 2 up to 80 Hz depending on the position in the region with SOs in the resistive part of the dependence $I_L(I_{t1})$. Outside the mentioned parts of the dependence $I_L(I_{t1})$ SOs of the current were absent.

Distinctive features of the dependences $I_L(I_{t1})$ are a value and a period of the current steps as well as a form of nonlinearities observed in the regions 2–7 and 5–8.

Turning to discussion of the dependences shown in Fig. 2, we highlight their three features: the critical state of PC1, corresponding to $I_{t1}=I_{c1}$ at which the current $I_L$ appears in the long branch of the DCS circuit; the equidistant current steps at $I_{c1}<I_{t1}<I_{c1}+I_{c2}$; the self-oscillations of the current $I_L$ in the resistive state of the both PCs.

The first two features have previously been found by us in Ref. 5. They are a part of more complex dependences shown in Fig. 2. Here we shall take a look at them in order to gain a deep insight into the physical processes, causing their appearance, and into the consequences following from their existence.

There two possible mechanisms causing the critical state of PC1 and leading to the current in the long branch. Let us consider the first of them. At $I_{t1} = I_{c1}$ the shunting of the contact by superconducting inductor $L$ for a direct current must not result in the appearance of its resistivity but it does not exclude a possible increasing of its Josephson parametric inductance $L_J$ as the current $I_{t1}$ approaches the value $I_{c1}$. In the case of a tunneling contact with the critical current $I_{c1}$ it is described by the formula,[6]

$$L_J = \Phi_0/(2\pi I_{c1} \cos \varphi), \quad (1)$$

where $\Phi_0$ is the magnetic flux quantum, $\varphi$ is the phase difference of wave functions of the Cooper pairs on the contact, which is a function of the transport current through the contact. Assuming that the PC is similar in its Josephson properties to a tunneling contact, the formula (1), as a first approximation, can be applied to the PC under investigation. It is seen that the theoretical value of $L_J$ can reach even infinity (at $\varphi = \pi/2$) and can be all the more considerably larger than $L = 5 \cdot 10^{-6}$ H resulting in the redistribution of the transport current $I_{t1}$ between the branches and in the appearance of the current in the much longer branch of the DCS. This mechanism of the practically dissipationless state of the PC differs from the traditional one applicable to an autonomous contact, not included to the DCS circuit when its critical current is determined by appearance of resistivity found from a current-voltage characteristic. In spite of the fact that the concept of the dissipative critical state of a superconducting Josephson contact has been known for quite long time from the theoretical works on the current distribution in quantum interference devices,[7] our result contains a new element. It consists in the following. Previously, using the other technique[8] only small (by a few percent) increase of $L_J$ in the tunneling Josephson contact has been demonstrated as the current through it approaches the critical value. From our measurements, if the assumption about a dissipationless mechanism of formation of critical state in the PC is correct, it follows that when the first step of the current $I_L$ appears the parametric inductance $L_J$ attains values greater than $5 \cdot 10^{-6}$ H. In this case the estimate of the classical inductance of the contact,[9] made on a base of its proposed geometric dimensions, gives the value of $10^{-11}$–$10^{-12}$ H. Thus, the inductance increases by more than a factor of $10^6$.

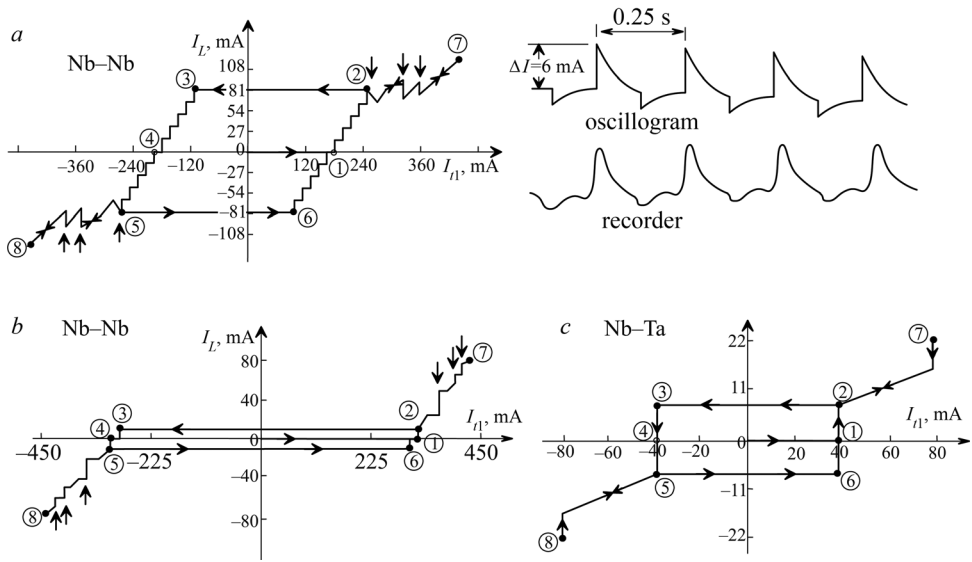

FIG. 2. Three main types of dependences of the current $I_L$ in the long branch of the DCS on the transport current $I_{t1}$, injected into one of the PCs at a different ratio between critical current of the PC in the long branch and the height of the current step, equal to 5 (a), 1 (b), <1 (c). The inset in (a) on the right-hand side shows a form of typical self-oscillations in branches of the DCS when both PCs are in the resistive state. The vertical arrows on the dependences denote the regions where self-oscillations of the current $I_L$ exist.

Let us consider the second formation mechanism of the critical state in the PC connected to the DCS circuit. It assumes a possibility of achieving the resistivity of the PC1 at $I_{t1} = I_{c1}$, although it is shunted by the superconducting inductor $L$ of the long branch of the DCS. At saturation of the surrounding environment with electromagnetic fields of different strength and with different frequencies there is practically no way to avoid random alternative currents induced by these fields in the circuit of the transport current and in the DCS circuit itself, in spite of the known effort taken to minimize them (filtering, screening, self-compensation). This can lead to the situation when even at the value of the direct transport current $I_{t1}$ close to $I_{c1}$, but still less than $I_{c1}$, it can be combined with half-waves of alternating currents coinciding in direction. As a result the total current can exceed the critical current $I_{c1}$, and the electromotive force induced in the inductor $L$ can compensate the voltage appearing on the PC. This will indicate the appearance of resistivity of the PC which can lead to increasing its temperature, to decreasing $I_{c1}$ and, as a consequence, to supplying the direct transport current to the long branch of the DCS. Upon realization of this mechanism the effect of the inductance $L_J$ on the formation of the critical state of the PC in the DCS can either be reduced or, in general, turned to zero depending on level of parasitic electromagnetic field acting on the DCS and its circuits.

The comparison of the considered mechanisms of formation of the critical state in the PC as a part of the DCS, and the experimental results obtained do not allow currently to support one of them. It indicates a strong demand for developing a strict theory of phenomena in this DCS, in spite of the complexity of processes taking place in them, and the necessity of further improving the experimental technique in this area.

The presence in Figs. 2(a) and 2(b) of the equidistant current steps can indicate, as assumed in Ref. 5, that the PC, through which the transport current is injected, has a microstructure of the superconducting interference device. For the first time the structure of the clamping contact-interferometer has been described in Ref. 10. It consists in several microcontacts in parallel with different critical currents, which randomly formed upon compression of two superconductors. A distinctive property of an interferometer with such a structure, as our experience indicates, can be violation of a strict periodicity of the voltage, appearing in it, in the resistive state during variation of an external magnetic field, typical of interferometres with two contacts. To confirm it we carried out a special experiment in one of the PCs, temporarily isolated form the DCS circuit which is characterized by the dependence $I_L(I_{t1})$ shown in Fig. 2(a). In this experiment the direct transport current exceeding the critical value passed through the PC, and the magnetic field $H$ was created in the intersection plane of niobium microwires by means of a special coil with a current. Fig. 3 shows the obtained dependence of the voltage $V$ in this PC on the magnetic field $H$, confirming the complex microstructure of this PC in form of the interferometer with several microcontacts in parallel. If we introduce the concept of the averaged period of this dependence over the field, which is taken to be equal to $\Delta H = 0.2$ Oe, then the quantization area $S_0$ of some equivalent two-contact interferometer can be estimated using the known relation,

$$S_0 = \Phi_0/(\mu_0 \kappa \Delta H), \qquad (2)$$

where $\mu_0 = 4\pi \cdot 10^{-7}$ H/m, and $\kappa$ is the concentration of magnetic field in the region of the interferometer located between two superconducting microwires with relatively big diameters. After substituting the numerical values, we obtain (at $\kappa = 10$) $S_0 = 10^{-11}$ m$^2$ (i.e., 10 $\mu^2$). Thus, an order of dimensions of the micro quantum interferometer appearing upon contact of niobium microwires was determined. This estimate can be used as a basis for developing the DCS with a two-contact interferometer (which is much simpler to analyze but more complicated technologically) which has the mentioned quantization area. Results of investigation of such a DCS can accelerate developing of a strict theory of the phenomena observed in these DCSs.

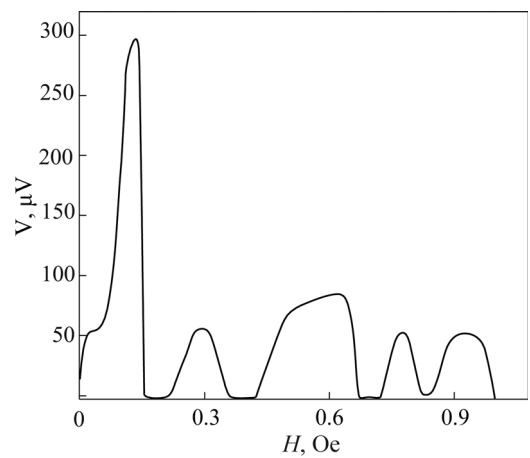

FIG. 3. Aperiodic dependence of the voltage across the PC on an external magnetic field the vector of which $H$ lies in the plane of intersection of microwires forming the point contact.

On the other hand, the experimental evidence of existence of the PS in form of an interferometer with a complex structure of microcontacts leads to the following question. Why does this structure not cause disturbances of the periodicity and the form of current steps in the dependence $I_L(I_{t1})$ as the current $I_{t1}$ increases, and thus of the magnetic field of the current acting on the interferometer to the moment when both contacts of the DCS reach the resistive state? Conversely, the periodicity of steps states that upon injection of such a transport current through one of multi-connected PCs the reaction of the DCS on it looks as if the contact was a two-contact interferometer, the model of which has been proposed by us in Ref. 5. This paradox can be explained if it is assumed that the periodicity of the current steps is achieved due to the presence of some threshold mechanism providing the step-like appearance of the current $I_L$ in the long branch of the DCS. In this case the strict periodicity of response of the input device on the input signal (in our case, the response of autonomous interferometer on varying magnetic field) is not needed. It can be assumed that such a mechanism is realized in this DCS with a clamping PC. It can appear as the following. At some current value $I_{t1} = I_{c1}$ the parametric inductance of microcontacts of the interferometer PC achieves the critical value (or the resistivity of the PC appears if the second mechanism of formation of the critical state occurs). The current quantum $I_L$ enters the superconducting long branch of the DCS, at the same time causing the decrease of the parametric induction of the interferometer (or disappearance of its resistivity in the case of the second mechanism) due to a feedback via a magnetic field of the current, and is frozen in the DCS circuit. The further increase of the current $I_{t1}$ through the interferometer PC and of the magnetic field created by it causes the growth in it of the circulating current up to the value when the critical state of the interferometer and the increased value of the parametric inductance of the PC, which corresponds to it, (or until the resistivity of the interferometer is reached) are achieved. After that a new jump of the current $I_L$ will occur. The value of a quantum of the current $I_{t1}$ and a corresponding quantum of the magnetic field of this current, determining the periodicity of the process and modulation of the critical current of the interferometer, is determined by all parameters of this multi-connected interferometer. Upon further increasing the current $I_{t1}$ the process repeats. A strict quantitative evidence of the proposed model of the current quantization process in the DCS with the PC requires developing and investigating a similar DCS with two-contact interferometer with parameters specified in advance, that is planned. From the given model of processes in the DCS it also follows that the distinction between three types of the dependences $I_L(I_{t1})$, shown in Fig. 2, in the region of their hysteresis parts can be explained by different relations between critical currents of the contacts (PC1 and PC2) and the depth of the quantum modulation of the critical current of interferometers formed by it.

Finally, let us consider the current distribution in the DCS after the resistive state is reached in both PCs. The appearance of some node point 2 in the dependences $I_L(I_{t1})$ (see. Fig. 2), where their slope changes and the hysteresis typical of them disappears, is explained as achieving by the current $I_{t1}$ (and hence by the current $I_L$) the value of the critical current $I_{c2}$ in the second PC (PC2), through which the transport current is not injected. Starting from this value of the $I_{t1}$ its distribution over the branches of the DCS is determined by the law different from previous one. It starts to depend in the resistive branches with the PC. The detailed study of the law of this distribution is still necessary to perform, in particular, after making a DCS with two-contact interferometers. At the same time, at this stage of studying the behavior of the current in the DCS with resistive PCs, the existence of SOs of the current in its branches at fixed values of the current $I_{t1}$ can be established. Before discussing a mechanism of the appearance of these SOs, it is necessary to note that different types of current and voltage self-oscillations in structures with superconductors fed by a direct current have been investigated for more than half a century.[11–24] The types can be divided into two groups. To the first belong SOs with specified frequency and amplitude,[11–22] to the second—chaotic and, as a rule, low-frequency SOs, amplitude and frequency of which vary in time randomly.[23,24] The current self-oscillations in our experiments belong to the first group. Let us dwell on the known structures where the SOs of this type were observed. Such structures can be called as structures of the L–R–S type, a close electric circuit of which contains a known inductance L, a normal specified resistance R, and also a superconducting element S, which is a point[13,18] or tunneling[14,19] Josephson contact or a film bridge.[16,17,20] The mentioned circuit is usually fed either from a direct current or constant voltage supply. The main condition of appearance of SOs in the mentioned structures is a hysteresis current-voltage characteristics (CVC) of the S-element. In particular, as the current exceeds a critical value there appear a jump of voltage on it and a sharp increase of its resistance up to some value $R^*$. As the current decreases, the hysteresis of voltage and the transition into the superconducting state at smaller critical current is observed. In the circuit L–R–S of the structure where the current is supplied from the voltage supply with an internal resistance $R_0 \ll R^*$, upon achievement of a critical current there appear SOs of the current with the frequency close to $f = (R + R^*)/L$ in the circuit. The difference between the structure of our DCS and that of previously known is that its closed circuit, consisting of inductor L and two PCs, does not contain the normal resistance $R$. Instead there are two resistivities $R_1^*$ and $R_2^*$, depending on current and its magnetic field, of point contacts which, as shown above, are quantum interferometers in the resistive state. The diagram of such a circuit, where for the sake of simplicity PC1 and PC2 are represented by two-contact interferometers with the quatization areas $S_{01}$ and $S_{02}$, is shown in Fig. 4. Besides the transport current, its magnetic field ($H_1^*$ and $H_2^*$) also acts on interferometers. The current source used in our experiments sets the voltage on interferometers when the relation $I_{t1} > I_{c1} + I_{c2}$ is satisfied. Assuming a quite probable difference in their resistivity (for instance, $R_1^* \ll R_2^*$), the interferometer with smaller resistivity becomes a source of voltage for a rest part of the DCS circuit, consisting of the inductance and the second interferometer with higher resistivity. In accordance with the structure of the L–R–S type described above and with the condition of appearing the current SOs in it, our structure in the resistive sate can be considered as one variety. The $L$ and $R_2^*$ in series are fed by the current from the voltage source with an internal resistance $R_1^*$. The difference is that the resistance of the $S$-element depends not only on the current passing through it but also on the magnetic field of this current ($H_1^*$ and $H_2^*$). Let us assume that in the beginning this current does not cause its self-oscillations, and we will increase it. In so doing, the magnetic field, created by it and acting on interferometers, changes. At some current a quite sharp increase of the magnetoresistance of the second interferometer can appear similar to that which exists at some values of the field $H$ in the dependence shown in Fig. 3. It leads to decreasing the current in the branch with the $L$ and $R_2^*$ in series. In turn, the current decreasing leads to decreasing the magnetic field in the region of the second interferometer and to corresponding back decreasing the resistance. Then the process will be repeated with the frequency close to $f_2 = R_2^*/L$ at the constant transport current $I_{t1}$. It is a possible mechanism of appearance of the current SOs with rather determined frequency in the DCS circuit with two resistive PCs. In order to estimate values of the frequency of SOs expected according to the supposed model it is necessary to know the value of $R_2^*$ at $I_{t1}$ in the vicinity of the critical current of the DCS. To determine it we use a CVC of one of the PCs reported in Ref. 5, which is similar in its properties to that considered in this work. In the vicinity of the critical current the differential resistance of this contact is $R_2^* \approx 5 \cdot 10^{-5}$ Ohm, that corresponds to the oscillation frequency of the current $f_2 = 10$ Hz. It corresponds, in order of magnitude, to experimental data. The absence of the current SOs at some intermediate values of the current $I_{t1}$ can also be due to the complex dependence of the voltage on the interferometers on the magnetic field of this current. In particular, for parts of the dependence of the voltage across the interferometer on the magnetic field shown in Fig. 3, where the voltage and, thus, the resistance of the contact depends weakly on the magnetic field, the current SOs must be absent or they are hard to find. On the whole, one can conclude that the proposed model of appearance of SOs corresponds qualitatively to real processes occurring in the resistive DCs with two clamping contacts.

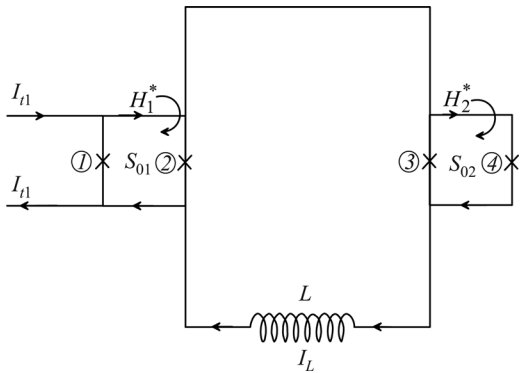

FIG. 4. A schematic diagram of two PCs in parallel in form of two-contact interferometers with quatization areas $S_{01}$ and $S_{02}$ and contacts 1, 2, 3, 4, the resistive state of which changes upon action of the current $I_{t1}$ and its magnetic field $H_1^*$ and $H_2^*$.

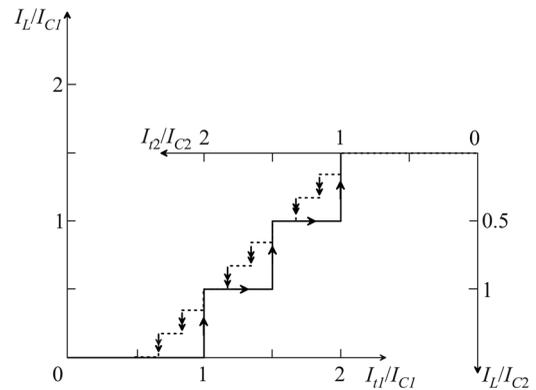

FIG. 5. Two conjugated dependences of a relative value of the current in the DCS circuit on a relative value of the transport current through the first and the second point contacts with current steps different in value, which demonstrate a possibility of its increasing by using the current $I_{t1}$ (solid line with arrows) and its subsequent decreasing by using the current $I_{t2}$ (dashed line with arrows).

Besides the experiments described above with injection of the current ($I_{t1}$) through one of the PCs, an effect on the dependence $I_L(I_{t1})$ of simultaneous injection of the current into the DCS from two current sources connected to the PC1 and PC2 were investigated. If the value and the direction of transport currents across them is chosen in such way that $I_{c1} < I_{t1} < I_{c2}$ and $I_{c2} < I_{t2} < I_{c1}$, then a step-like increase of the current $I_L$ up to required values is possible with using the current $I_{t1}$ without the current $I_{t2}$. After that at the fixed value of the current $I_{t1}$, the current $I_{t2}$ can be increased up to the value higher than $I_{c2}$, and the decrease of the current $I_L$ previously reached with periodicity of steps typical of the contact interferometer PC2 can be obtained. An illustration of the adjustment of the current $I_L$ is shown in Fig. 5 as dependences $I_L/I_{c1}$ ($I_{t1}/I_{c1}$) and $I_L/I_{c2}$ ($I_{t2}/I_{c2}$) superimposed for the sake of comparison of the adjustment processes. Here, the steps on the dependences $I_L/I_{c1}$ and $I_L/I_{c2}$ are different due to the formation process of structure of interferometers by the contacts PC1 and PC2, which is random in its origin. So using both transport currents leads to freezing the current $I_L$, which was achieved before turning off in the DCS circuit. Thus, the other way (besides reported in Ref. 5) to increase and decrease the current $I_L$ with its subsequent freezing by a fine adjusting of the relation between the currents $I_{t1}$ and $I_{t2}$ was found.

6## CONCLUSION

A microstructure of modern high-temperature ceramic and granular superconductors as well as multi-cored superconducting cables represents a multi-connected medium. The current self-oscillations found for the first time in a doubly connected superconductor with two resistive point contacts, as one unit cell of such a medium, can cause generation of noise magnetic field and fluctuations of resistance in these superconductors when they are in the resistive state. Furthermore, it is of interest to verify realization peculiarities of the proposed mechanism of appearing the self-oscillations in doubly connected superconductors with other types of weak links which do not have such a complex and nonrepetitive substructure as in a clamping PC, in particular in film DCSs with bridge contacts.

Regarding the solution of the question about a mechanism of the current critical state of a PC shunted by a superconducting inductor, the most preferable mechanism is dissipation-less and related to increasing the parametric Josephson inductance of the contact as the current through it approaches a critical value. A strict evidence of this assumption is possible, providing works in two directions: via deeper experimental verification of influence of surrounding electromagnetic fields on the DCSs and via developing a quantitative theory explaining the step-like dependence of the current in the DCS on the transport current through it.

For application, new method of increasing and decreasing the current in the superconducting circuit of the DCS by the transport current, injected through two contacts simultaneously, seems to be more convenient in comparison with the adjustment methods proposed in Ref. 5 for superconducting magnets. Furthermore this method can be used for designing multi-terminal superconducting memory devices on the base of the DCS containing either two or more inputs of tuning transport current in form of a PC.

[1] A. Rose-Innes and E. Roderick, *Vvedenie v fiziku sverhprovodimosti (Introduction to Physics of Superconductivity)* (Mir, Moscow, 1972).
[2] *Slabaya sverhprovodimost, Kvantovye interferometry i ih primenenie (Weak Superconductivity, Quantum Interferometers and their Application)*, edited by B. B. Schwartz and S. Fonner (Mir, Moscow, 1980).
[3] M. Tinkham, *Vvedenie v sverhprovodimost (Introduction to superconductivity)* (Atomizdat, Moscow, 1980).
[4] S. I. Bondarenko, V. P. Koverya, A. V. Krevsun, N. M. Levchenko, and A. A. Shablo, Fiz. Nizk. Temp. **36**, 202 (2010) [Low Temp. Phys. **36**, 159 (2010)].
[5] V. P. Koverya, S. I. Bondarenko, A. V. Krevsun, N. M. Levchenko, and I. S. Bondarenko, Fiz. Nizk. Temp. **36**, 759 (2010) [Low Temp. Phys. **36**, 605 (2010)].
[6] K. K. Likharev and B. T. Ulrikh, *Sistema s dzhozevsonovskimi kontaktami (System with Josephson contacts)* (Izd-vo Moskovskogo universiteta, Moscow, 1978).
[7] W.-T. Tsang and T. Van Duzer, J. Appl. Phys. **46**, 4573 (1975).
[8] A. H. Silver, R. C. Jaklevic, and J. Lambe, Phys. Rev. **141**, 362 (1966).
[9] P. L. Kalantarov and L. A. Zeitlin, *Raschet induktivnostey (Calculation of Inductors)* (Energoatomizdat, Leningrad, 1986).
[10] A. H. Silver and J. E. Zimmerman, Phys. Rev. **157**, 317 (1967).
[11] G. B. Rosenberger, IBM J. Res. Dev. **3**, 189 (1959).
[12] S. Y. Berkovitch, Radiotehnika i elektronika (Radiotechnique and Electronics) **4**, 736 (1965).
[13] J. E. Zimmerman and A. H. Silver, Phys. Rev. Lett. **19**, 14 (1967).
[14] F. L. Vernon, Jr. and R. J. Pedersen, J. Appl. Phys. **19**, 2661 (1968).
[15] I. I. Eru, S. A. Peskovatskii, and A. V. Poladich, Fiz. Tverd. Tela **15**, 2228 (1973).
[16] H. Frohlich, H. Koch, W. Vodel, D. Wachter, and O. Frauberger, Physica **64**, 197 (1973).
[17] W. J. Skocpol, M. R. Beasley, and M. Tinkham, J. Appl. Phys. **45**, 4054 (1974).
[18] Y. Taur and P. L. Richards, J. Appl. Phys. **48**, 1793 (1975).
[19] N. Calander, T. Cleason, and S. Rudner, Appl. Phys. Lett. **39**, 504 (1981).
[20] M. Muck, H. Rogalla, and C. Heiden, Appl. Phys. A **46**, 97 (1988).
[21] K. Enpuku, T. Kisu, and K. Yoshida, IEEE Trans. Magn. **27**, 3058 (1991).
[22] V. A. Rakhubovskii, Voprosy atomnoy nauki i tehniki (Prob. At. Sci. Technol.), No. 6, 105 (2009).
[23] A. V. Khotkevich and I. K. Yanson, Fiz. Nizk. Temp. **7**, 69 (1981) [Sov. J. Low Temp. Phys. **7**, 35 (1981)].
[24] V. P. Koverda, V. N. Skokov, and V. P. Skripov, Pis'ma v Zh. Eksp. Teor. Fiz. **63**, 739 (1996).